\documentclass[journal]{IEEEtran}
\usepackage{amsmath,amssymb,amsfonts,mathrsfs, mathtools, bm, bbm, dsfont, mathrsfs,epstopdf, amsthm}
\usepackage{cite}
\usepackage{graphicx}
\usepackage[dvipsnames]{xcolor}

\usepackage{stfloats,subfigure,multicol,epsfig,epsf,psfrag,latexsym,graphicx}
\usepackage{chngpage}
\usepackage{textcomp}

\usepackage{url}
\usepackage{array}

\usepackage{hyperref}
\usepackage{breakurl}

 \def\old#1{}    

\def\nn{\nonumber}
\def\beq{\begin{equation}}
\def\eeq{\end{equation}}
\def\bea{\begin{eqnarray}}
\def\eea{\end{eqnarray}}
\def\ba{\begin{array}}
\def\ea{\end{array}}

\def\bitem{\begin{itemize}}
\def\eitem{\end{itemize}}
\def\ben{\begin{enumerate}}
\def\een{\end{enumerate}}

\def\ie{{\it i.e.,\ \/}}

\def\alphabf{\hbox{\boldmath$\alpha$\unboldmath}}

\def\epsilonbf{\hbox{\boldmath$\epsilon$\unboldmath}}

\def\etabf{\hbox{\boldmath$\eta$\unboldmath}}

\def\lambdabf{\hbox{\boldmath$\lambda$\unboldmath}}
\def\mubf{\hbox{\boldmath$\mu$\unboldmath}}

\def\xibf{\hbox{\boldmath$\xi$\unboldmath}}
\def\pibf{\hbox{\boldmath$\pi$\unboldmath}}

\def\omegabf{\hbox{\boldmath$\omega$\unboldmath}}

\def\dbf{{\bm d}}

\def\fbf{{\bm f}}

\def\pbf{{\bm p}}

\def\ubf{{\bm u}}
\def\vbf{{\bm v}}

\def\Abf{{\bm A}}

\def\Cbf{{\bm C}}

\def\Pbf{{\bm P}}

\def\Rbf{{\bm R}}
\def\Sbf{{\bm S}}

\def\Xbf{{\bm X}}

\newcommand{\beqa}{\begin{eqnarray}}
\newcommand{\eeqa}{\end{eqnarray}}
\newcommand{\beqan}{\begin{eqnarray*}}
\newcommand{\eeqan}{\end{eqnarray*}}

\newcommand\T{{\mathpalette\raiseT\intercal}}
\newcommand\raiseT[2]{\raisebox{0.25ex}{$#1#2$}
}

\newcommand{\Rset}{\mathbb{R}}

\newcommand{\Scal}{{\cal S}}

\newcommand{\argmax}{\mathop{\rm argmax}}

\renewcommand{\v}[1]{{\bm{#1}}}

\newcommand{\ol}[1]{\ensuremath{\overline{{#1}}}}
\newcommand{\ul}[1]{\ensuremath{\underline{{#1}}}}

\newcounter{l1}
\newcounter{l2}
\newcounter{l3}
\newcommand{\bdotlist}{\begin{list}{$\bullet$}{}}
\newcommand{\bboxlist}{\begin{list}{$\Box$}{}}
\newcommand{\bbboxlist}{\begin{list}{\raisebox{.005in}{{\tiny
$\blacksquare$ \ \ }}}{}}
\newcommand{\bdashlist}{\begin{list}{$-$}{} }
\newcommand{\blist}{\begin{list}{}{} }
\newcommand{\barablist}{\begin{list}{\arabic{l1}}{\usecounter{l1}}}
\newcommand{\balphlist}{\begin{list}{(\alph{l2})}{\usecounter{l2}}}
\newcommand{\bAlphlist}{\begin{list}{\Alph{l2}.}{\usecounter{l2}}}
\newcommand{\bdiamlist}{\begin{list}{$\diamond$}{}}
\newcommand{\bromalist}{\begin{list}{(\roman{l3})}{\usecounter{l3}}}

\newtheorem{proposition}{Proposition}

\newtheorem{definition}{Definition}

\title{DSO-DERA Coordination for the Wholesale Market Participation of Distributed Energy Resources}

\author{Cong Chen,
        ~Subhonmesh Bose,
        ~and Lang~Tong
\thanks{\scriptsize
Cong Chen and Lang Tong (\{cc2662, lt35\}@cornell.edu) are with the School of Electrical and Computer Engineering, Cornell University, Ithaca NY, USA. Subhonmesh Bose (\url{boses@illinois.edu}) is with the Department of Electrical and Computer Engineering at the University of Illinois Urbana-Champaign (UIUC), Urbana IL, USA.}
}

\begin{document}
\maketitle

\begin{abstract}
We design a coordination mechanism between a distribution system operator (DSO) and distributed energy resource aggregators (DERAs) participating directly in the wholesale electricity market. Aimed at ensuring system reliability while providing open access to DERAs, the coordination mechanism includes a forward auction that allocates access limits to aggregators based on aggregators' bids that represent their benefits of aggregation. The proposed coordination mechanism results in decoupled DSO and DERAs operations that satisfy the network constraints, independent of the stochasticity of the renewables, wholesale real-time locational marginal prices, and individual DERA's aggregation policy. Optimal bidding strategies by competitive aggregators are also derived. The efficiency of the coordination mechanism and the locational price distribution at buses of a radial distribution grid are demonstrated for a 141-bus radial network.

\end{abstract}

\begin{IEEEkeywords}
DER aggregation, behind-the-meter distributed generation, network access allocation mechanism.
\end{IEEEkeywords}

\section{Introduction}\label{sec:Intro}

The landmark ruling of the Federal Energy Regulatory Commission (FERC) Order 2222 aims to remove barriers to the direct  participation of distributed energy resource aggregators (DERAs) in the wholesale market operated by Regional Transmission Organizations and Independent System Operators (RTO/ISO) \cite{FERC20}. Since DERs are situated in the distribution systems, services procured from aggregated DERs must pass through the distribution grid operated by a distribution utility or a distribution system operator (DSO\footnote{Herein, we assume that the DSO is either the utility or an independent entity that operates the distribution system.}).  To this end, a DSO-DERA coordination mechanism is necessary to ensure system reliability on the one hand and open access for all DERAs on the other. FERC Order 2222 recognizes the significance of DSO-DERA coordination but leaves the specifics of the coordination to the utility, DERAs, and the regulators.   

DSO-DERA coordination poses significant theoretical, practical, and economic challenges. Power injections and withdrawals from DERs will likely depend on the wholesale market condition (such as locational marginal prices and real-time needs for regulation services) and the available resources (e.g.,  behind-the-meter renewables). Yet, with DER and wholesale price uncertainties, the DSO must ensure the safe operation of the distribution grid to reliably deliver power to all customers of the distribution utility and the DERAs.
Any coordination mechanism that the DSO adopts must provide open and equitable access to multiple competing DERAs operating over the same distribution network. By equitable access in this context, we mean that the  mechanism cannot discriminate among the DERAs in how they participate or are compensated. 


DSO-DERA-ISO coordination is being actively debated since the release of FERC Order 2222. In \cite{Renjit:22EPRIReport},  coordination models have been classified into four categories, ranging from the least to the most DSO involvement. Type I models assume no DSO control (e.g., see \cite{Alshehri&etal:20TPS,Gao&Alshehri&Birge:22,chen22competitive}), and installed DER capacities are deemed to lie within the network's hosting capacity limits. As a result, system reliability is guaranteed for arbitrary power injection/withdrawal profiles from DERs. Type II models explicitly consider the randomness of power injections from DERs, where the DSO strives to prevent constraint violations, e.g., see \cite{NazirAlmassalkhi22TPS}. Type III models involve coordination between the DSO and the ISO, where DERAs can also provide distribution grid services besides wholesale market products. Type IV models require DERA aggregation through the DSO, with DSO performing all reliability functions and participating in the wholesale market on behalf of DERAs  as in \cite{GirigoudarRoald22T64, chen2019aggregate, MousaviWu22TPS}.

\begin{figure}[htbp]
    \centering
    \includegraphics[scale=0.8]{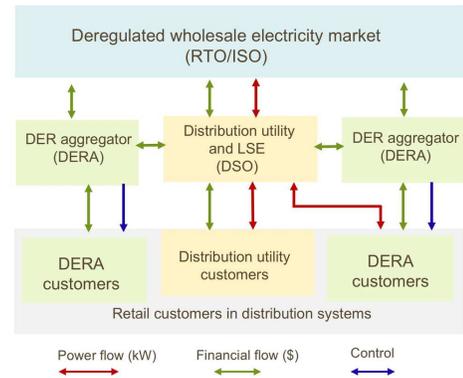}
    \caption{Power flow, financial flow and control interactions in the  DERA-DSO coordination model.}
    \label{fig:DERAmodel}
\end{figure}

We propose a type-II DSO-DERA coordination mechanism under the generic interaction model among DSO, RTO/ISO, and multiple DERAs, as shown in Fig. 1 from \cite{chen22competitive}, where DERAs and DSO operate {\em independently} under clearly defined operational limits to ensure system reliability.   In particular, the access limit of each DERA is allocated through a forward auction of available network capacities based on DERAs' bids and offers ahead of the actual operation. The forward auction present in Section \ref{sec:AccessRight} is constructed such that system operation constraints are not violated as long as each DERA aggregates within its cleared access limits. Each DERA then exercises its aggregation and wholesale market participation policies subject to the allocated access limits. We formulate a DERA's profit maximization problem from DER aggregation in Section \ref{sec:InjDERAopt} and compute a DERA's optimal bid in the access limit allocation problem.
There is no need for DSO intervention in the ISO dispatch of a DERA in our model as long as the  DER injections from DERA customers respect the access limit that the DERA procures from the DSO. Thus, our mechanism entails ``minimal'' DSO-DERA interaction based on physical network constraints and DERAs' economic incentives. Furthermore, the  DSO does not need to know DERAs' aggregation strategies and customer preferences. Likewise, when designing and executing an aggregation policy, a DERA does not need to be aware of the network constraints and other DERAs' models. Despite the minimal interaction between the DSO and the DERAs, our design guarantees distribution system constraint enforcement. In addition, numerical experiments in Section \ref{sec:CaseStudies} reveal that the social surplus of the market increases with more customers switching from incumbent distribution utilities to DERAs, making our design aligned with the spirit of FERC Order 2222.

We remark that our proposed DSO-DERA coordination mechanism may appear similar to the transmission right allocation problem for participants with bilateral contracts considered in the early years of wholesale market deregulation. Allocating physical transmission rights was deemed impractical nor necessary (see \cite{Hogan92JRE_FTR}) with the ``loop-flow'' problem in meshed transmission networks. Ultimately, wholesale markets evolved to adopt a centrally coordinated economic dispatch run by the ISO, where bilateral transactions came to be hedged using financial transmission rights--an electricity derivative. In our coordination mechanism, on the other hand, the access limit allocation is physical, allowing DERAs to inject or withdraw any amount of power within their purchased access limits. The loop-flow problem does not affect our design, thanks to the radial nature of distribution networks. Also, we deliberately separate the access limit allocation from dispatch decisions; we envision the auction for access limits to run only once every day or week. We posit that tight coordination of dispatch decisions via a centralized market mechanism matched to that operated by the ISO may be impractical in the near term and possibly unnecessary, owing to smaller trade volumes and less stringent system constraints in the distribution grid.

\section{DSO's Access Allocation Mechanism}\label{sec:AccessRight}
We consider a mechanism where the DSO auctions off accesses across the distribution network to the DERAs.

Consider $K$ DERAs operating over a distribution network across $N$ buses. Let the power injection profile from DERA $k$ across the network be given by $\v{p}_k \in \Rset^{N}$. Thus, the collection of DERAs inject $\v{p} := \sum_{k=1}^K \v{p}_k$ amount of real power. Let the power injection profile from the utility's customers be given by $\v{p}_0 \in \Rset^N$. Assuming a uniform power factor for all power injections, the reactive power injection profile is given by $\alpha (\v{p}+\v{p}_0)$. These real and reactive power injections then induce power flows and voltage magnitudes over the distribution network that are related via the power flow equations. In this work, we adopt a linear distribution power flow (LinDistFlow) model  \cite{baranWu89TPDdistFlow} that relates the power flows $\v{f} \in \Rset^{N-1}$ over the distribution lines and the voltage magnitudes $\v{v} \in \Rset^{N-1}$ via 
\begin{align}
\begin{pmatrix}
\v{f} \\ \v{v} \end{pmatrix} 
 = \begin{pmatrix} 
 \v{A}_{fp}  & \v{A}_{fq}
 \\
 \v{A}_{vp} & \v{A}_{vq}
 \end{pmatrix} 
 \begin{pmatrix} \v{p} + \v{p}_0 \\ \alpha (\v{p} + \v{p}_0)\end{pmatrix} 
 = \v{A}(\v{p} + \v{p}_0).
\end{align}
The derivation of $\v{A}$ for the LinDistFlow model is included in the appendix. The power flows and the voltage magnitudes must remain within the engineering limits of the network, as 
\begin{align}\label{eq:volLBUB}
   \v{\ul{b}}^\T := (\v{\ul{f}}^\T, \v{\ul{v}}^\T)^\T \leq (\v{f}^\T, \v{v}^\T)^\T \leq (\v{\ol{f}}^\T, \v{\ol{v}}^\T)^\T :=  \v{\ol{b}}^\T.
\end{align}

With this power flow model in hand, we now describe the network access auction mechanism that the DSO runs. Notice that we expect that the auction will take place once every day or week. As a result, the DSO must account for a variety of operating conditions over the horizon so that the results of the auction remain binding. Let DERA $k$ provide a bid, $\varphi_k:\Rset^{N\times N} \rightarrow \Rset$, to acquire the access to an injection capacity $\v{\ul{C}}_k\in \Rset^{N}$ and a withdrawal capacity $\v{\ol{C}}_k\in \Rset^{N}$ with the understanding that all power injections from assets controlled by DERA $k$ must respect
$    \v{p}_k \in [\v{\ol{C}}_k, \v{\ul{C}}_k]
$. Define $\v{\ol{C}}:=(\v{\ol{C}}_k),\v{\ul{C}}:=(\v{\ul{C}}_k)$ as the matrices collecting access limits across DERAs. Let $(\v{\ol{C}}_k^{\mbox{\tiny max}}, \v{\ul{C}}_k^{\mbox{\tiny max}})$ be the maximum injection and withdrawal capacities that DERA $k$ decides to purchase. Let the utility's own customers have net power injections $\v{p}_0$ that take values in the set $[\v{\ul{p}}_0, \v{\ol{p}}_0]$. Given these ranges of the various power injections, the DSO solves the following robust optimization to clear the forward auction allocating accesses for the distribution network capacities to each DERA.
\begin{align}
    \begin{aligned}
    & \underset{\v{\ul{C}}, \v{\ol{C}}, \v{\ol{P}}, \v{\ul{P}}}{\text{maximize}} && \sum_{k=1}^K \varphi_k(\v{\ul{C}}_k,\v{\ol{C}}_k)-J(\ol{\Pbf},\ul{\Pbf}), 
    \\
    & \text{such that} 
    &&{\bm 0} \le \v{\ol{C}}_k \le \v{\ol{C}}_k^{\mbox{\tiny max}},
    \\
    &&& \v{\ul{C}}_k^{\mbox{\tiny max}} \le \v{\ul{C}}_k \le {\bm 0},
    \\
    &&& \v{\ol{P}} = \sum_{k=1}^K \v{\ol{C}}_k + \v{\ol{p}}_0,
    \\
    &&&
    \v{\ul{P}} = \sum_{k=1}^K \v{\ul{C}}_k + \v{\ul{p}}_0,
    \\ 
    &&& 
    \v{p} =\sum_{k=1}^K \v{p}_k,
    \\
    &&&
    \v{\ul{b}}
    \leq \v{A}(\v{p} + \v{p}_0) 
    \leq 
    \v{\ol{b}},
    \\
    &&& \text{for all } \v{p}_k \in [\v{\ul{C}}_k, \v{\ol{C}}_k], \v{p}_0 \in [\v{\ul{p}}_0, \v{\ol{p}}_0].
    \end{aligned}
    \label{eq:auction}
\end{align}
Here, the functions $\varphi_k$ are the induced utilities from DERA $k$. The offer formation is described in the next section. $\ol{\Pbf}$ and $\ul{\Pbf}$ respectively represent the total injection and withdrawal capacity at each bus. And the function, $J:\Rset^{N\times N} \rightarrow \Rset$,  is the operation cost of DSO including reactive power support, network maintenance, etc. The problem in \eqref{eq:auction} enforces the engineering constraints of the grid for every possible power injection profile from the utility's customers and those of all DERAs within their acquired capacities. This problem is a robust linear program with a continuum of constraints. However, the rectangular nature of the constraint sets for the variables $\v{p}_k$'s and $\v{p}_0$ allow an easy reformulation. We rewrite $\v{A} = \v{A}_+ - \v{A}_-$, where the entries of the matrices on the right-hand side are constructed out of the positive and negative parts of the respective entries of $\v{A}$.
\begin{align}
    \begin{aligned}
    & \underset{\v{\ul{C}}, \v{\ol{C}}, \v{\ol{P}}, \v{\ul{P}}}{\text{maximize}} && \sum_{k=1}^K\varphi_k(\v{\ul{C}}_k,\v{\ol{C}}_k)-J(\ol{\Pbf},\ul{\Pbf}), 
    \\
    & \text{such that} 
    &&{\bm 0} \le \v{\ol{C}}_k \le \v{\ol{C}}_k^{\mbox{\tiny max}},
    \\
    &&& \v{\ul{C}}_k^{\mbox{\tiny max}} \le \v{\ul{C}}_k \le {\bm 0},
    \\
    &~~~~~~~~\ol{\lambdabf}:&& \v{\ol{P}} = \sum_{k=1}^K \v{\ol{C}}_k + \v{\ol{p}}_0,
    \\
    &~~~~~~~~\ul{\lambdabf}:&&
    \v{\ul{P}} = \sum_{k=1}^K \v{\ul{C}}_k + \v{\ul{p}}_0,
    \\
    &~~~~~~~~\ol{\mubf}:&&
    \v{A}_+ \v{\ol{P}} - \v{A}_-  \v{\ul{P}}
    \leq 
    \v{\ol{b}},
    \\
    &~~~~~~~~\ul{\mubf}:&&
    \v{A}_+  \v{\ul{P}} - \v{A}_- \v{\ol{P}} \geq 
    \v{\ul{b}}.
    \end{aligned}
    \label{eq:auction.2}
\end{align}
Here, the objective function represents the social surplus, \ie
\beq
{\Scal}(\v{\ul{C}}, \v{\ol{C}}, \v{\ol{P}}, \v{\ul{P}}):=\sum_{k=1}^K\varphi_k(\v{\ul{C}}_k,\v{\ol{C}}_k)-J(\ol{\Pbf},\ul{\Pbf}).
\eeq
 which is the surplus of all DERAs minus the operation cost of DSO for a certain injection/withdrawal access allocation.
Associate Lagrange multipliers with the constraints as indicated above. 
We define the locational allocation price as follows.

\begin{definition}[Locational Allocation  Prices]
Optimal dual solutions $\ol{\lambdabf}^\star \in \Rset^N, \ul{\lambdabf}^\star \in \Rset^N$ of (\ref{eq:auction.2}), respectively, define the vector of locational injection and withdrawal allocation prices for DERAs at different buses.
\end{definition}
These prices are uniform at each distribution bus and hence, do not discriminate among DERAs that operate at the same bus. The prices, however, may differ with location in the distribution network. Given these prices, DERA $k$ with DER allocations $\ol{C}_k^i, \ul{C}_k^i$ at bus $i$ pays the DSO $\ol{C}_k^i\ol{\lambda}_i^\star-\ul{C}_k^i\ul{\lambda}_i^\star$.

One can view our auction pricing as a special case of marginal pricing as follows. To that end, let $\ol{\epsilonbf}$ denote the injection access imbalance, given by $\v{\ol{P}}-(\sum_{k=1}^K \v{\ol{C}}_k + \v{\ol{p}}_0) = \ol{\epsilonbf}$. (Similarly, the withdrawal access imbalance $\ul{\epsilonbf}$.) Then, consider \eqref{eq:auction.2} with the $\v{\epsilon}$-perturbed access imbalance equations, and its optimal social surplus ${\cal S}^*(\ol{\epsilonbf},\ul{\epsilonbf})$. With this notation, we have the following result.

\begin{proposition}\label{Prop:LAP}
The locational allocation prices satisfy
\begin{align*}
    \ol{\lambdabf}^\star 
    = \nabla_{\ol{\epsilonbf}}{\cal S}^\star(0,0) 
    = \nabla_{\ol{\Pbf}}J(\ol{\Pbf}^\star,\ul{\Pbf}^\star)+\Abf_+^{\T}\ol{\mubf}^\star - \Abf_-^{\T}\ul{\mubf}^\star,
    \\
    \ul{\lambdabf}^\star
    = \nabla_{\ul{\epsilonbf}}{\cal S}^\star(0,0)=\nabla_{\ul{\Pbf}}J(\ol{\Pbf}^\star,\ul{\Pbf}^\star)-\Abf_-^{\T}\ol{\mubf}^\star+\Abf_+^{\T}\ul{\mubf}^\star.
\end{align*}
\end{proposition}
Proof of the above proposition follows the envelope theorem for the first equal sign and KKT conditions of \eqref{eq:auction.2} for the second equal sign.  Details are illustrated in the appendix. Assume that DSOs adopt operation cost $J$ that is uniform across all buses, and DERAs are homogeneous over different buses. Then, if no line capacity and voltage constraints bind at an optimal solution of \eqref{eq:auction.2}, the access prices become uniform across the network.

\section{Bids and Offers from Profit-seeking DERAs }\label{sec:InjDERAopt}
We now explain the offer formation for the induced benefit functions $\varphi_k$ of DERA $k$. All variables and parameters in this section are associate with DERA $k$, and we omit subscript $k$ for simplicity. DERAs construct their bids for the network access auction by maximizing their profits assuming forecast available renewable. 

We first extend the DER aggregation approach in \cite{chen22competitive} to the case with injection access constraints for the optimal decision-making of DERA with the aggregated prosumers. With this injection access-constrained optimization, we compute  the optimal bids $\varphi$ from DERA to the forward auction (\ref{eq:auction.2}) allocating network accesses. 
 
\subsection{Profit Maximizing DER Aggregation}
 
We consider the case that DERA maximizes her own surplus, while providing $\zeta>1 $ times those surpluses to the aggregated customers she serves than the utility's rate for distributed generation (DG). This section is an extension of the DER aggregation approach in \cite{chen22competitive} to the case with injection access constraints. Without loss of generality, we assume each bus has one prosumer, thus the DERA aggregating $N$  prosumers will need to have injection/withdrawal accesses for these $N$ buses. DERA receives  payment $\omega^i \in \Rset$ from each prosumer indexed by $i$ and directly buys or sells its aggregated energy in the wholesale market with the locational marginal price $\pi_{\mbox{\tiny LMP}} \in \Rset$. To maximize DERA surplus between the received payment and the payment to the wholesale market,  DERA uses the optimization below. 
\begin{align}
    \begin{aligned}
    \varphi(\v{\ul{C}},\v{\ol{C}}) =& \underset{\omegabf,\dbf}{\text{maximum}} && 
    \sum_{i=1}^N (\omega^i-\pi_{\mbox{\tiny LMP}} (d^i-r^i)), 
    \\
    & \text{such that}  
    &&\zeta S^i_{{\textrm{NEM}}}(r^i) \le U^i(d^i)-\omega^i,
    \\
    &&&
    \ul{d}^i \le d^i \le  \ol{d}^i,
    \\ 
    &&& 
    \ul{C}^{i} \le   r^i-d^i \le  \ol{C}^{i},
    \\
    &&& i =1,\ldots,N.   
    \end{aligned}
    \label{eq:DERAsurplus_LnGP}
\end{align}

The first constraint encodes the  competitive nature of DERA, which  guarantees that the prosumer surplus is at least $\zeta$ times the surplus under the net energy metering (NEM X). Explicit formulation of prosumer surplus $S^i_{{\textrm{NEM}}}(r^i)$ is shown in the appendix. The second constraint is prosumer consumption limits, and the last constraint is injection and  withdrawal access limits.

\subsection{Benefits of Aggregation and Bid Formation}
The following proposition introduces the optimal closed-form solution of (\ref{eq:DERAsurplus_LnGP}). The optimal DERA surplus is decomposed into three terms, the surplus dependent on the withdraw access $\ul{\varphi}^i({\ul{C}}^i)$, the surplus dependent on the  injection access $\ul{\varphi}^i({\ul{C}}^i)$, and the surplus independent of the network access $h^i-\zeta S^i_{{\textrm{NEM}}}(r^i)$. 
 
\begin{proposition}[Optimal DERA Decision]\label{Prop:DERA}
The optimal solution of \eqref{eq:DERAsurplus_LnGP} is given by
\begin{align}
\begin{aligned}
d^{i\star}(\pi_{\mbox{\tiny LMP}},r^i)
&=\min\{r^i-\ul{C}^{i}, \max\{\hat{d}^i, r^i-\ol{C}^{i}\}\},
\\
\omega^{i\star}(d^{i\star},r^i)
&= U^i(d^{i\star})-\zeta S^i_{{\textrm{NEM}}}(r^i),
\end{aligned}
\end{align}
where $\hat{d}^i:=\min\{{\ol{d}}^i,\max\{(V^i)^{-1}(\pi_{\mbox{\tiny LMP}}), {\ul{d}}^i\}\}$.

The optimal DERA surplus is given by
\begin{align}\label{eq:bidCurve}
\varphi(\v{\ul{C}},\v{\ol{C}})=&\sum_{i=1}^N\ul{\varphi}^i({\ul{C}}^i)+\sum_{i=1}^N\ol{\varphi}^i({\ol{C}}^i)\\\nn
&+\sum_{i=1}^N (h^i-\zeta S^i_{{\textrm{NEM}}}(r^i)),
\end{align}

\begin{align*}
\text{where } h^i&:=U^i(\hat{d}^i)-\pi_{\mbox{\tiny LMP}} (\hat{d}^i-r^i),\\
\ul{\varphi}^i({\ul{C}}^i)&:=\begin{cases}
U^i(r^i-\ul{C}^{i})+\pi_{\mbox{\tiny LMP}}\ul{C}^{i}-h^i, & \text{if } r^i\le   q_+^i,\\
0,&\text{otherwise},
\end{cases}
\\
\ol{\varphi}^i({\ol{C}}^i)&:=\begin{cases}
U^i(r^i-\ol{C}^{i})+\pi_{\mbox{\tiny LMP}}\ol{C}^{i}- h^i, & \text{if }q_-^i \le r^i,\\
0,& \text{otherwise},
\end{cases}
\\
q_-^i &:=\ol{C}^{i} +\max\{(V^i)^{-1}(\pi_{\mbox{\tiny LMP}}),\ul{d}^i\},
\\
q_+^i &:=\ul{C}^{i}+\min \{(V^i)^{-1}(\pi_{\mbox{\tiny LMP}}),\ol{d}^i\}.
\end{align*}
\end{proposition}
The proof follows the Karush-Kuhn-Tucker optimality conditions for \eqref{eq:DERAsurplus_LnGP} (see appendix for details). When there's no binding constraints for network access limits, $\hat{d}^i$ is the optimal consumption of prosumer $i$, and $h^i-\zeta S^i_{{\textrm{NEM}}}(r^i)$ is the benefit of DERA from prosumer $i$. When there's binding network access constraints, the optimal consumption $d^{i\star}(\pi_{\mbox{\tiny LMP}},r^i)$ equals to $\hat{d}^i$ truncated by the network access limits, and benefit of DERA is modified by $\ol{\varphi}^i({\ol{C}}^i)$ and $\ul{\varphi}^i({\ul{C}}^i)$. 

We remark that the optimal DERA benefit $\varphi(\v{\ul{C}},\v{\ol{C}})$ is separable across injection and withdrawal access, and across prosumers at different distribution buses. Furthermore, at most one of $\ol{\varphi}^i({\ol{C}}^i)$ and $\ul{\varphi}^i({\ul{C}}^i)$ can be nonzero, depending upon the renewable generation $r^i$ for prosumer $i$. When $r^i\le   q_+^i$, $\ul{\varphi}^i({\ul{C}}^i)$ is nonzero and prosumer at bus $i$ is a consumer with binding network withdrawal access constraints; when $q_-^i \le  r^i $, $\ol{\varphi}^i({\ol{C}}^i)$ is nonzero and prosumer at bus $i$ is a producer with binding network injection access constraints.  

Therefore, DERA can construct its bids for injection and withdrawal access as  $\ol{\varphi}^i({\ol{C}}^i)$ and $\ul{\varphi}^i({\ul{C}}^i)$ at different buses. In practice, piece-wise linear apprixmations of those bids are needed to be included in the access auction problem (\ref{eq:auction.2}).  
\section{Numerical Experiments}\label{sec:CaseStudies}
We considered a 141-bus radial distribution network with 4 DERAs who aggregate households at different buses with different levels of behind-the-meter distributed generation (BTM DG). 
Parameters for resistances and reactances of this system are taken from \cite{khodr08EPSR141,MatpowerCase141}. We assumed fixed power factor of 0.98 across all buses and set line flow limit of 15MW for the first 6 branches, consecutively connected to the substation bus, and 2MW for the rest. All 4 DERAs aggregate resources from all buses, except that DERA 4 only aggregates over buses 118-134. The BTM DG for households under DERA 1, 2, 3, and 4 are respectively, $0.2$kW, $5.2$kW, $10.2$kW, and $15.2$kW. We set maximum access limits for all DERAs,\ie $\ol{\Cbf}_k^{\mbox{\tiny max}}, \ul{\Cbf}_k^{\mbox{\tiny max}}$, equal to  $0.1$ MW. We adopted a quadratic cost for the DSO of the form $J(x)=\frac{1}{2}bx^2-ax$ with  $a=-0.096, b=0.2$.

We used homogeneous utility functions for   households  \cite{SamadiSchoberWong12TSG}, 
 $$U(x)=\begin{cases}
\hat{a} x-\frac{\hat{b}}{2}x^2,&0\le x\le \frac{\hat{a}}{\hat{b}},\\ 
\frac{\hat{a}^2}{2\hat{b}},&x> \frac{\hat{a}}{\hat{b}},
\end{cases}$$ 
with $\hat{a}=0.65, \hat{b}=0.2$. The percentage of DG adopters over all households is assumed $80\%$. For NEM X tariff, we used $\pi^0=\$0$, $\pi^+=\pi^{-}+\$0.03$/kWh, $\zeta=1.003$, and  the wholesale market LMP $\pi_{\mbox{\tiny  LMP}}=\$0.03$/kWh \cite{CAISO22price}. 

\begin{figure}[htbp]
    \centering
    \includegraphics[scale=0.47]{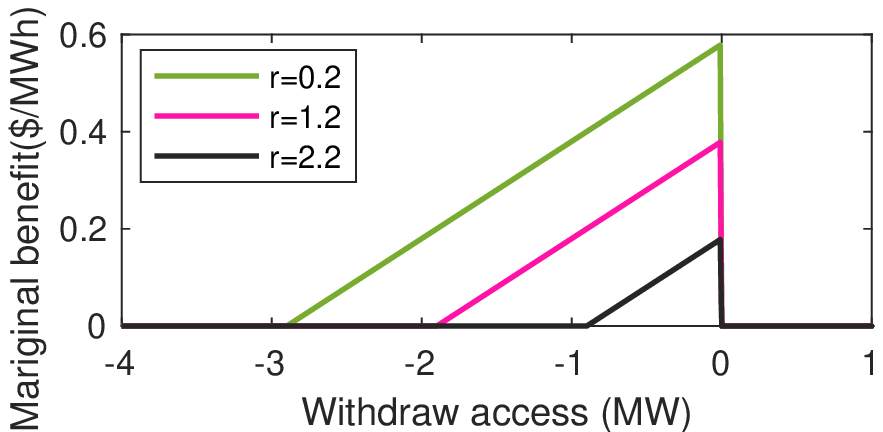}\includegraphics[scale=0.47]{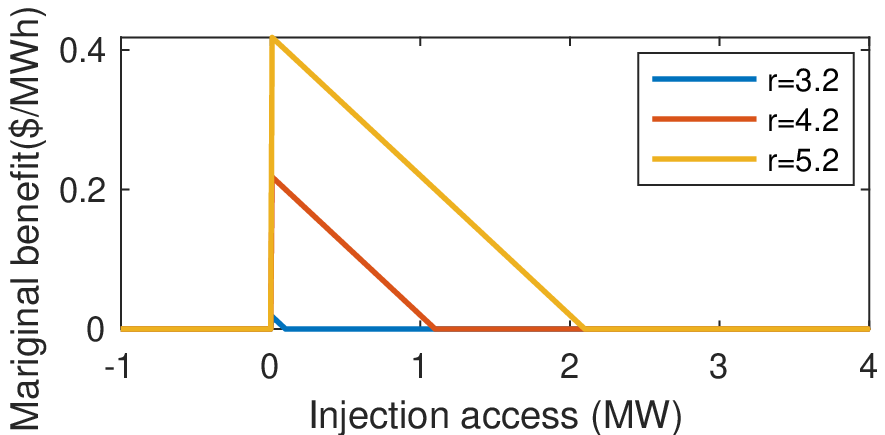}
    \caption{Left: Bid-in marginal benefit vs withdrawal access. Right: Bid-in marginal benefit vs injection  access.}
    \label{fig:INJ}
\end{figure}

The supply offers can be constructed from \eqref{eq:bidCurve}. We plot here instead the marginal benefits $\ul{\varphi}^{i,'}_k, \ol{\varphi}^{i,'}_k$ in Fig.~\ref{fig:INJ} with varying BTM DG generations. The left panel illustrates that  DERAs with lower DG levels choose higher bid prices to purchase withdrawal access. The right panel shows that DERAs with higher DG  choose higher prices for injection, rather than withdrawal, access.

\begin{figure}[htbp]
    \centering
    \includegraphics[scale=0.46]{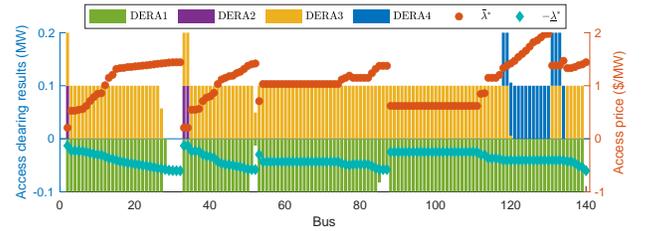}
    \caption{Market clearing results of  network access.}
    \label{fig:MC}
\end{figure}

Results of DSO's auction with the 4 DERAs is illustrated in Fig.~\ref{fig:MC}. The positive and negative segments of the left y-axis respectively represent the allocated injection and withdrawal ranges. DERA 1 has low DG levels, and behaves as a net consumer, who bids for and receives withdrawal access. Fig.~\ref{fig:MC} shows that DERA 1 received withdrawal security operating range $[0, 0.1]$ MW over most buses. DERAs 2, 3 and 4 largely act as net producers of power, and they therefore purchase injection access through the auction. A DERA with higher DG level has higher marginal surplus for injection access. One expects this higher marginal surplus to lead to higher clearing prices for injection. Not surprisingly, buses  118-134 with DERA 4 exhibit the highest levels of $\ol{\lambda}^i$. Also, binding line and voltage limits cause access prices to vary with location. We indeed observe this effect for $\ul{\lambda}^i, \ol{\lambda}^i$ across all buses consecutively connected to buses 28-32, 52, 86-87, 130, and 140-141, which have binding voltage constraints. Locational allocation prices are gradually increasing along the branch until reaching the bus with  binding voltage constraint. 



\begin{figure}[htbp]
    \centering
    \includegraphics[scale=0.5]{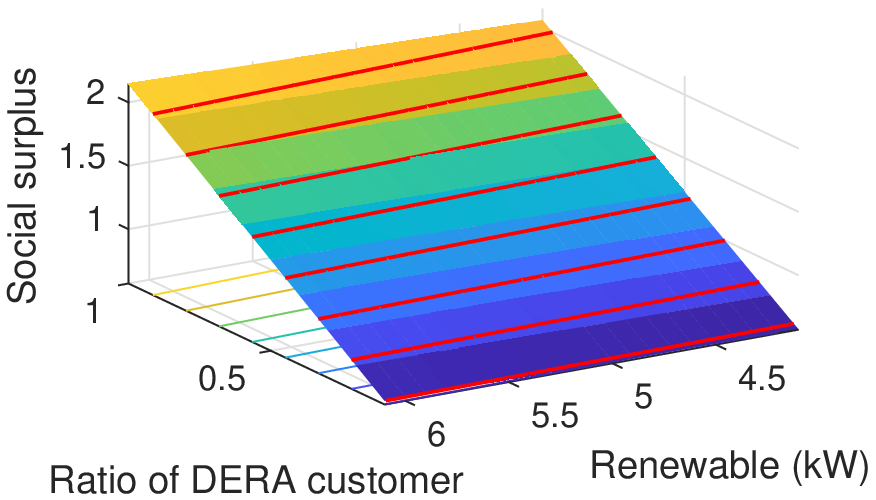}\includegraphics[scale=0.5]{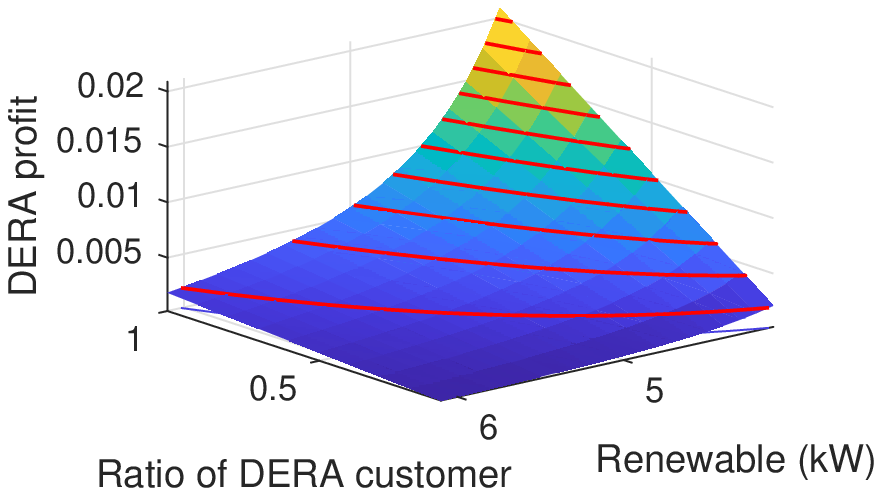}
\caption{Left: social surplus. Right: DERA surplus.}
    \label{fig:SurplusDERASocial}
\end{figure}

The left panel of Fig.~\ref{fig:SurplusDERASocial} shows a normalized version of the social surplus $\Scal$. Here, $\Scal$ increases, when the ratio of DERA customer increases, or when DG level increases. To uncover the reason behind such phenomenon, notice that the first constraint of DERA aggregation model \eqref{eq:DERAsurplus_LnGP} schedules its customers at a demand level with higher surplus compared to the distribution utility with NEM X tariff. With more customers switching from the distribution utility to DERAs, the total social surplus for in the DSO's auction  increases.

The right panel of Fig.~\ref{fig:SurplusDERASocial} reveals that the total surplus of all DERAs decreases with increasing DG, it increases with the ratio of DERA customer base. Since NEM X provides higher surplus to customers with higher DG, as noted in \cite{AlahmedTong2022NEM}, DERAs must commensurately reduce their own profits and share them with the customers to remain competitive with NEM X.






\section{Conclusions}\label{sec:Conclusion}
This paper proposes a distribution network injection access allocation mechanism for DSO to allocate its network injection access to multiple DERA participants. Such a mechanism offers a simple decomposition for DERA-DSO coordination. DSO can guarantee voltage and line flow security for the distribution network when the DERAs guarantee that power injections respect DSO's injection access allocation. Such an allocation mechanism runs once every day or week, and gives more freedom to  DERAs to coordinate the decentralized aggregation of prosumers' assets. 

We also derived optimal bid curves for DERAs to participate in the forward auction for distribution network accesses, when DERAs offer compensations that are competitive with respect to distribution tariff structures such as those for NEM. Through numerical experiments, we observed that DERAs with higher BTM generation generally bid higher for injection access than those with lower generation. DERAs with lower BTM generation tend to bid higher for withdrawal access. We also observed that. the social surplus of the retail market increases with more customers switching from incumbent DSO utility companies to DERAs.


{
\bibliographystyle{IEEEtran}
\bibliography{BIB}
}

\section{Appendix}\label{sec:Appendix}
 \subsection{Prosumer Under NEM X Tariff}

We explain the optimal decision-making for prosumers with the behind-the-meter distributed generation (BTM DG) under net energy metering (NEM X) from the incumbent utility company. More details and variants of the optimal prosumer decisions are explained in \cite{chen22competitive}.

 Consider a prosumer with  energy consumption $d \in \Rset_+$  and  renewable generation $r \in \Rset_+$. Assume that the prosumer's utility, $U:\Rset_+ \rightarrow \Rset$,  is strictly concave, monotonically increasing, and continuously differentiable with  marginal utility function $V:=U'$ invertible over $[0, +\infty)$. Prosumer index $i$ is dropped here for simplicity.
 
We assume that prosumer consumption is independent of the DG output $r$, i.e., all generation is used for bill reductions. The NEM-X tariff is utilized for the retail rate, which is parameterized by $\pibf=(\pi^+, \pi^-, \pi^0)$ as in \cite{AlahmedTong2022NEM}. $\pi^+ \in \Rset$ is the retail (consumption) rate, $\pi^- \in \Rset$ is the sell (production) rate, and $\pi^0 \in \Rset$ is the connection charge. The optimal consumption of this prosumer is given by
\begin{align}\label{eq:passiveD}
\begin{aligned}
    d_{{\textrm{NEM}}} 
&= \argmax_{ \ul{d} \leq d \leq \ol{d}} \left(U(d)-\pi^+d\right).
\end{aligned}
\end{align}\vspace{-0.1cm}

So, we have $ d_{{\textrm{NEM}}} =\min\{{\ol{d}},\max\{V^{-1}(\pi^+), {\ul{d}}\}\}$. The optimal prosumer surplus  is given by
\begin{align}
    &S_{{\textrm{NEM}}}(r)=\begin{cases}
U(d_{{\textrm{NEM}}} )-\pi^-(d_{{\textrm{NEM}}} -r)-\pi^0,& \text{if }r \geq d_{{\textrm{NEM}}} \nn,\\
U(d_{{\textrm{NEM}}} )-\pi^+(d_{{\textrm{NEM}}} -r)-\pi^0,& \text{otherwise}. 
\end{cases}
\end{align}

\newpage

\subsection{Proof of Proposition \ref{Prop:LAP}}
We use the envelope theorem to prove the first equal sign and use KKT conditions of \eqref{eq:auction.2} for the second equal sign.

Consider \eqref{eq:auction.2} with the $\v{\epsilon}$-perturbed access imbalance equations, the optimal social surplus ${\cal S}^\star(\ol{\epsilonbf},\ul{\epsilonbf})$ can be computed by
\begin{align}
    \begin{aligned}
    {\cal S}^\star(\ol{\epsilonbf},\ul{\epsilonbf})=& \underset{\v{\ul{C}}, \v{\ol{C}}, \v{\ol{P}}, \v{\ul{P}}}{\text{maximize}} && \sum_{k=1}^K\varphi_k(\v{\ul{C}}_k,\v{\ol{C}}_k)-J(\ol{\Pbf},\ul{\Pbf}), 
    \\
    & \text{such that} &&\\
    &~~~(\ul{\etabf},\ol{\etabf}):&&{\bm 0} \le \v{\ol{C}}_k \le \v{\ol{C}}_k^{\mbox{\tiny max}},
    \\
    &~~~(\ul{\xibf},\ol{\xibf}):&& \v{\ul{C}}_k^{\mbox{\tiny max}} \le \v{\ul{C}}_k \le {\bm 0},
    \\
    &~~~~~~~~\ol{\lambdabf}:&& \v{\ol{P}}-(\sum_{k=1}^K \v{\ol{C}}_k + \v{\ol{p}}_0) = \ol{\epsilonbf},
    \\
    &~~~~~~~~\ul{\lambdabf}:&&
    \v{\ul{P}} - (\sum_{k=1}^K \v{\ul{C}}_k + \v{\ul{p}}_0)= \ul{\epsilonbf},
    \\
    &~~~~~~~~\ol{\mubf}:&&
    \v{A}_+ \v{\ol{P}} - \v{A}_-  \v{\ul{P}}
    \leq 
    \v{\ol{b}},
    \\
    &~~~~~~~~\ul{\mubf}:&&
    \v{A}_+  \v{\ul{P}} - \v{A}_- \v{\ol{P}} \geq 
    \v{\ul{b}}.
    \end{aligned}
    \label{eq:auction.epsilon}
\end{align}

The Lagrangian function of (\ref{eq:auction.epsilon}) is
\beq
\begin{array}{cc}
&    {\cal L}(\v{\ul{C}}, \v{\ol{C}}, \v{\ol{P}}, \v{\ul{P}}, \ol{\lambdabf}, \ul{\lambdabf}, \ol{\mubf}, \ul{\mubf})= J(\ol{\Pbf},\ul{\Pbf})-\sum_{k=1}^K\varphi_k(\v{\ul{C}}_k,\v{\ol{C}}_k) \\
 &~~  +\ol{\lambdabf}^\T( \ol{\epsilonbf}-(\v{\ol{P}}-(\sum_{k=1}^K \v{\ol{C}}_k + \v{\ol{p}}_0))) +\ul{\etabf}^\T(-\v{\ol{C}}_k)\\
 &~~+\ol{\etabf}^\T(\v{\ol{C}}_k - \v{\ol{C}}_k^{\mbox{\tiny max}}) +\ul{\lambdabf}^\T( \ul{\epsilonbf}-(\v{\ul{P}}-(\sum_{k=1}^K \v{\ul{C}}_k + \v{\ul{p}}_0)))\\
&~~    +\ol{\mubf}^\T(\v{A}_+ \v{\ol{P}} - \v{A}_-  \v{\ul{P}} -  \v{\ol{b}})
    +\ul{\mubf}^\T(\v{A}_+  \v{\ul{P}} - \v{A}_- \v{\ol{P}}- \v{\ul{b}})\\
    &~~+\ul{\xibf}^\T( \v{\ul{C}}_k^{\mbox{\tiny max}}-\v{\ul{C}}_k)+\ul{\xibf}^\T(\v{\ul{C}}_k).
\end{array}
\eeq

By envelope theorem $ \nabla_{\ol{\epsilonbf}}{\cal L}^\star 
    = \nabla_{\ol{\epsilonbf}}{\cal S}^\star , \nabla_{\ul{\epsilonbf}}{\cal L} ^\star
    = \nabla_{\ul{\epsilonbf}}{\cal S}^\star $. And we know that $\nabla_{\ol{\epsilonbf}}{\cal L}^\star=\ol{\lambdabf}^\star, \nabla_{\ul{\epsilonbf}}{\cal L}^\star=\ul{\lambdabf}^\star$. The locational allocation price for DERA injection is solved at $\ol{\epsilonbf}={\bm 0}, \ul{\epsilonbf}={\bm 0}$. So, we have $\ol{\lambdabf}^\star 
    = \nabla_{\ol{\epsilonbf}}{\cal S}^\star(0,0), \ul{\lambdabf}^\star
    = \nabla_{\ul{\epsilonbf}}{\cal S}^\star(0,0) $.
    
KKT conditions of \eqref{eq:auction.2}  gives
\beq
\begin{array}{cc}
\nabla_{\ol{\Pbf}}{\cal L}^\star =\nabla_{\ol{\Pbf}}J(\ol{\Pbf}^\star,\ul{\Pbf}^\star)-\ol{\lambdabf}^\star+\Abf_+^{\T}\ol{\mubf}^\star - \Abf_-^{\T}\ul{\mubf}^\star={\bm 0},\\
\nabla_{\ul{\Pbf}}{\cal L}^\star =\nabla_{\ul{\Pbf}}J(\ol{\Pbf}^\star,\ul{\Pbf}^\star)-\ul{\lambdabf}^\star-\Abf_-^{\T}\ol{\mubf}^\star+\Abf_+^{\T}\ul{\mubf}^\star={\bm 0}.
\end{array}
\eeq
So, we have $\ol{\lambdabf}^\star=\nabla_{\ol{\Pbf}}J(\ol{\Pbf}^\star,\ul{\Pbf}^\star)+\Abf_+^{\T}\ol{\mubf}^\star - \Abf_-^{\T}\ul{\mubf}^\star={\bm 0}$, and $\ul{\lambdabf}^\star=\nabla_{\ul{\Pbf}}J(\ol{\Pbf}^\star,\ul{\Pbf}^\star)-\Abf_-^{\T}\ol{\mubf}^\star+\Abf_+^{\T}\ul{\mubf}^\star$.

\subsection{Proof of Proposition \ref{Prop:DERA}}
We prove this proposition with KKT conditions of (\ref{eq:DERAsurplus_LnGP}), and the inventory calculation with the optimal solution. 

Assign dual variables to (\ref{eq:DERAsurplus_LnGP}), we have
\begin{align}
    \begin{aligned}
    \varphi(\v{\ul{C}},\v{\ol{C}}) =& \underset{\omegabf,\dbf}{\text{maximum}} && 
    \sum_{i=1}^N (\omega^i-\pi_{\mbox{\tiny LMP}} (d^i-r^i)), 
    \\
    & \text{such that} &&\\ 
    &~~~~~~~~~\chi^i:&&\zeta S^i_{{\textrm{NEM}}}(r^i) \le U^i(d^i)-\omega^i,
    \\
    &~~~~~(\ul{\nu}^i, \ol{\nu}^i):&&
    \ul{d}^i \le d^i \le  \ol{d}^i,
    \\ 
    &~~~~~(\ul{\gamma}^i, \ol{\gamma}^i):&& 
    \ul{C}^{i} \le   r^i-d^i \le  \ol{C}^{i},
    \\
    &&& i =1,\ldots,N.   
    \end{aligned}
    \label{eq:DERAsurplus_dual}
\end{align}
From KKT conditions of (\ref{eq:DERAsurplus_dual}), we have $\forall i=1,...,N$
\bea\label{eq:DERA_KKT}
\begin{array}{lrl}
\frac{\partial \cal L}{\partial \omega^i}=\chi^{i\star}-1=0,\\
\frac{\partial \cal L}{\partial d^i}=\pi_{\mbox{\tiny LMP}}-\chi^{i\star}V^i(d^{i*})-\ul{\nu}^{i\star}+\ol{\nu}^{i\star} +\ul{\gamma}^{i\star}-\ol{\gamma}^{i\star} =0.
\end{array}
\eea

Combined with the complementary slackness condition, the first constraint of (\ref{eq:DERAsurplus_dual}) is always binding with $\chi^{i\star}=1$, and the optimal consumption $d^{i\star}$ equals to $(V^i)^{-1}(\pi_{\mbox{\tiny LMP}})$ if it falls into the interval of the minimum upper bound $\min\{\ol{d}^i, r^i-\ul{C}^{i}\}$ and maximum lower bound $\max\{\ul{d}^i, r^i-\ol{C}^{i}\}$. So we have

\begin{align}
d^{i\star}(\pi_{\mbox{\tiny LMP}},r^i)
&=\min\{r^i-\ul{C}^{i}, \max\{\hat{d}^i, r^i-\ol{C}^{i}\}\},
\end{align}
\begin{align}
    \omega^{i\star}(d^{i\star},r^i)&= U^i(d^{i\star})-\zeta S^i_{{\textrm{NEM}}}(r^i),
\end{align}
where  $\hat{d}^i:=\min\{{\ol{d}}^i,\max\{(V^i)^{-1}(\pi_{\mbox{\tiny LMP}}), {\ul{d}}^i\}\}$.

\underline{When $(V^i)^{-1}(\pi_{\mbox{\tiny LMP}}) \geq \min\{\ol{d}^i, r^i-\ul{C}^{i}\}=r^i-\ul{C}^{i}$}, we have  $r^i\le  \ul{C}^{i} +\min\{\ol{d}^i, (V^i)^{-1}(\pi_{\mbox{\tiny LMP}})\}$. The optimal value can be computed by
\begin{align}
    \begin{aligned}
    &\varphi(\v{\ul{C}},\v{\ol{C}})=\sum_{i=1}^N (\omega^{i\star}-\pi_{\mbox{\tiny LMP}} (d^{i\star}-r^i))\\
    &~~~~=\sum_{i=1}^N(U^i(r^i-\ul{C}^{i})-\zeta S^i_{{\textrm{NEM}}}(r^i)-\pi_{\mbox{\tiny LMP}} (r^i-\ul{C}^{i}-r^i))\\
    &~~~~=\sum_{i=1}^N(U^i(r^i-\ul{C}^{i})+\pi_{\mbox{\tiny LMP}} \ul{C}^{i}-\zeta S^i_{{\textrm{NEM}}}(r^i)).
    \end{aligned}
\end{align}

\underline{When $(V^i)^{-1}(\pi_{\mbox{\tiny LMP}}) \le \max\{\ul{d}^i, r^i-\ol{C}^{i}\}=r^i-\ol{C}^{i}$}, we have  $r^i\geq  \ol{C}^{i} +\max\{\ul{d}^i, (V^i)^{-1}(\pi_{\mbox{\tiny LMP}})\}$. The optimal value can be computed by
\begin{align}
    \begin{aligned}
    &\varphi(\v{\ul{C}},\v{\ol{C}})=\sum_{i=1}^N (\omega^{i\star}-\pi_{\mbox{\tiny LMP}} (d^{i\star}-r^i))\\
    &~~~~=\sum_{i=1}^N(U^i(r^i-\ol{C}^{i})-\zeta S^i_{{\textrm{NEM}}}(r^i)-\pi_{\mbox{\tiny LMP}} (r^i-\ol{C}^{i}-r^i))\\
    &~~~~=\sum_{i=1}^N(U^i(r^i-\ol{C}^{i})+\pi_{\mbox{\tiny LMP}} \ol{C}^{i}-\zeta S^i_{{\textrm{NEM}}}(r^i)).
    \end{aligned}
\end{align}

\underline{In all other cases}, the optimal value is given by the equation below, which is not a function of $(\v{\ul{C}},\v{\ol{C}})$.
\begin{align}
    \begin{aligned}
    &\varphi=\sum_{i=1}^N(U^i(\hat{d}^i)-\zeta S^i_{{\textrm{NEM}}}(r^i)-\pi_{\mbox{\tiny LMP}} (\hat{d}^i-r^i)).
    \end{aligned}
\end{align}

Denote that $h^i:=U^i(\hat{d}^i)-\pi_{\mbox{\tiny LMP}} (\hat{d}^i-r^i)$,
\beq
\begin{array}{cc}
&\ul{\varphi}^i({\ul{C}}^i):=\begin{cases}
U^i(r^i-\ul{C}^{i})+\pi_{\mbox{\tiny LMP}}\ul{C}^{i}-h^i, & \text{if } r^i\le  q_+^i,\\
0,&\text{otherwise},\nn
\end{cases}
\end{array}
\eeq

\beq
\begin{array}{cc}
&\ol{\varphi}^i({\ol{C}}^i):=\begin{cases}
U^i(r^i-\ol{C}^{i})+\pi_{\mbox{\tiny LMP}}\ol{C}^{i}-h^i, & \text{if } q_-^i \le r^i,\\
0& \text{otherwise},\nn
\end{cases}
\end{array}
\eeq
 and $\begin{cases}q_-^i:=\ol{C}^{i} +\max\{(V^i)^{-1}(\pi_{\mbox{\tiny LMP}}),\ul{d}^i\},\\
q_+^i:= \ul{C}^{i} +\min \{(V^i)^{-1}(\pi_{\mbox{\tiny LMP}}),\ol{d}^i\}.\end{cases}$

To sum up over all cases, we have 
\beq
\begin{array}{clc}
\varphi(\v{\ul{C}},\v{\ol{C}})&=&\sum_{i=1}^N\ul{\varphi}^i({\ul{C}}^i)+\sum_{i=1}^N\ol{\varphi}^i({\ol{C}}^i)\\
&&+\sum_{i=1}^N (h^i-\zeta S^i_{{\textrm{NEM}}}(r^i)).
\end{array}
\eeq

\subsection{Example of linearized power flow}\label{sec:ExLPF}
In this section, we first describe parameters for the linearized distflow model  \cite{baranWu89TPDdistFlow}, and then provide a 5-bus example.
Consider a radial distribution networks with $N$ buses. Denote $u^i:=(v^i)^2, \forall i \in \{1,...,N\}$, and $u_{base}:=v_{base}^2$. Let the parent base bus (bus 1) be the reference bus, \ie $v^1=\sqrt{u_{base}}$. Consider voltage deviation no larger than 5\%. Then, with reduced shift factor matrix $\tilde{\Sbf}$, the bound for voltage in (\ref{eq:volLBUB}) is given by \footnote{Reduced shift factor matrix is shift factor matrix without the column for the reference bus.}
\beq\label{eq:volbd}
\begin{cases}
\ol{\vbf}:=1.05\ubf_{base}-\tilde{\Sbf}^\intercal(:,1)
u_{base},\\
\ul{\vbf}:=0.95\ubf_{base}-\tilde{\Sbf}^\intercal(:,1)u_{base},\end{cases}
\eeq
where we have $\ol{\v{v}}:=(\ol{v}^2, \ol{v}^3, \ol{v}^4, \ol{v}^5)^\intercal \in \Rset^{N-1}$ , and similarly $\ul{\v{v}} \in \Rset^{N-1}$ ignoring voltage at the reference bus.

Consider a fixed power factor in our model, we have reactive power injection $\alphabf \pbf=(\alpha_1 p_1,...,\alpha_n p_n,...,\alpha_N p_N)$ at different buses, given the active power $\pbf$. The forward auction for distribution network accesses in (\ref{eq:auction}) uses $\Abf $ represent the shift factor matrix for voltage and line flow in the linear power flow model. Detail parameter settings for $\Abf$ is listed below by
\beq
\Abf:=\left[
\begin{array}{c|c}
{\bf 0} & \tilde{\Sbf}\\
\hline
{\bf 0} & 2\tilde{\Sbf}^\intercal(\Rbf +\alpha \Xbf)
\end{array}
\right],
\eeq
where $\Rbf$ and $\Xbf$ are matrices for resistance and reactance. 

\begin{figure}[htbp]
    \centering
    \includegraphics[scale=0.45]{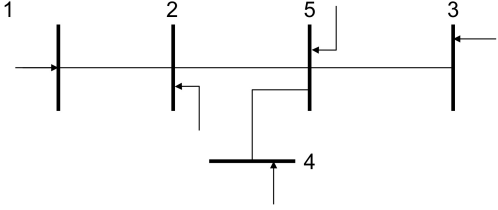}
    \caption{5-bus radial distribution network.}
    \label{fig:c}
\end{figure}

Consider a 5-bus system shown in the following figure. Let the parent base bus (bus 1) be the reference bus. With matrix rows consecutively related to line 2-1, line 5-2, line 3-5 and line 4-5, and matrix columns to bus 1, 2, 3, 4, 5, the shift factor matrix and reduced shift factor matrix are  $\Sbf=\begin{bmatrix}
0&1&1&1&1 \\
0&0&1&1&1\\
0&0&1&0&0\\
0&0&0&1&0
\end{bmatrix},~ \tilde{\Sbf}=\begin{bmatrix}
1&1&1&1 \\
0&1&1&1\\
0&1&0&0\\
0&0&1&0
\end{bmatrix}$, respectively.

The matrix for resistance and reactance are  respectively
$\Rbf=\begin{bmatrix}
r_{21}&r_{21}&r_{21}&r_{21}\\
0&r_{52}&r_{52} &r_{52} \\
0&r_{35}&0&0\\
0&0&r_{45}&0
\end{bmatrix},~ \Xbf=\begin{bmatrix}
x_{21}&x_{21}&x_{21}&x_{21}\\
0 &x_{52}&x_{52}&x_{52} \\
0&x_{35}&0&0\\
0&0&x_{45}&0
\end{bmatrix}$.

The power flow equation is

$$
2\left[
\begin{array}{c|c}
\Rbf & \Xbf\\
\end{array}
\right]\begin{bmatrix}
 \pbf_{\mbox{\tiny -1}}\\
 \alphabf \pbf_{\mbox{\tiny -1}}
\end{bmatrix}
=(\tilde{\Sbf}^\intercal)^{-1}
\ubf_{\mbox{\tiny -1}}-\begin{bmatrix}
u_{base}\\
{\bf 0}
\end{bmatrix}
$$
\beq\label{eq:pf}
\Rightarrow 2\tilde{\Sbf}^\intercal(\Rbf +\alpha \Xbf) \pbf_{\mbox{\tiny -1}}+\tilde{\Sbf}^\intercal(:,1)
u_{base}=\ubf_{\mbox{\tiny -1}},\eeq
where $\pbf_{\mbox{\tiny -1}}:=(p^2+p_0^2,p^3+p_0^3,p^4+p_0^4,p^5+p_0^5)^\intercal$, and $\ubf_{\mbox{\tiny -1}}:=(u^2,u^3,u^4,u^5)^\intercal$.




Denote  $\ol{\v{f}}:=(\ol{f}_{21}, \ol{f}_{52}, \ol{f}_{35}, \ol{f}_{45})^\intercal \in \Rset^{N-1}$ for branch flow limits (similarly $\ul{\v{f}} \in \Rset^{N-1}$), and combine (\ref{eq:volbd}) and (\ref{eq:pf}), we have the following equation.
\beq
\begin{array}{c}
\begin{bmatrix}\underline{\fbf}\\
\ul{\vbf}
\end{bmatrix}
\leq 
\left[
\begin{array}{c|c}
{\bf 0} & \tilde{\Sbf}\\
\hline
{\bf 0} & 2\tilde{\Sbf}^\intercal(\Rbf +\alpha \Xbf)
\end{array}
\right]
(\v{p} + \v{p}_0) 
 \leq    \begin{bmatrix}\bar{\fbf}\\
\ol{\vbf}
\end{bmatrix},
\end{array}
\eeq
which is the explicit form of 
the power flow constraint in (\ref{eq:auction}), \ie $\v{\ul{b}}
    \leq \v{A} (\v{p} + \v{p}_0) 
    \leq 
    \v{\ol{b}}$. 

\end{document}